\documentclass[vecphys]{svmult}

\usepackage{makeidx}         
\usepackage{graphicx}        
\usepackage{multicol}        
\usepackage[bottom]{footmisc}

\makeindex             

\begin{document}
\def\lsim{~\rlap{$<$}{\lower 1.0ex\hbox{$\sim$}}}
\def\gsim{~\rlap{$>$}{\lower 1.0ex\hbox{$\sim$}}}
\title*{Threshold Effects Beyond the Standard Model}
\author{Tomasz R. Taylor}
\institute{Department of Physics, Northeastern University, Boston, MA 02115, U.S.A.} 
%
%
\maketitle
I am very happy to contribute to the Festschrift celebrating Gabriele Veneziano on his 65th birthday. I have known Gabriele for more than 25 years, and worked with him on many projects, learning not only physics, but how to {\em en{\nolinebreak}joy} physics. ``Amusing'' is the word that he often uses to describe interesting ideas, and that single word characterizes best a unique style of {\em joyful} research that led to his pioneering work on string theory, particle physics and cosmology described in this book.
When researching Gabriele's original work on running coupling constants, preceding our 1988 collaboration \cite{Taylor:1988vt} described below, I ran into a write-up of his lectures on ``Topics in String Theory'' delivered in 1987 in China and in India \cite{Veneziano:1988aj}. His paper concludes with: ``But my moral, I hope, is a clear one for the young string theorist: If string math.\ is lots of fun, string phys.\ is no less.'' Indeed, I had much fun working on string physics over the following twenty years.
In this contribution, I discuss the threshold effects of extra dimensions and their applications to physics beyond the standard model, focusing on superstring theory.

\section{Threshold Effects from Extra Dimensions}
At a given time in the history of elementary  particle physics, there is always the mystery of  higher energies and the hope of building even more powerful accelerators that would take us one step farther in the understanding of short-distance physics. Thirty years ago, discovering yet another quark was a major breakthrough but now, the next round of experiments can hardly satisfy theorists without uncovering extra dimensions or producing black hole fireballs. Threshold effects appear each time a new particle is discovered. They appear in many physical quantities, signaling transition to new energy domains. 

As an example, consider the top quark threshold. We want to see how the QCD coupling constant evolves from the region below the top mass scale $m_t$, across the threshold, to higher energies. 
\begin{figure}\centering
\includegraphics[scale=0.6]{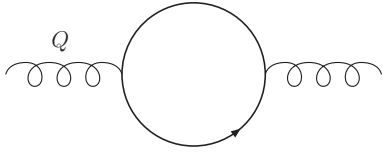}
\caption{One-loop contributions to the effective gauge coupling}
\end{figure}
In order to determine the corresponding one-loop correction to the effective action we can consider the vacuum polarization  diagram with two external gauge bosons at momentum scale $Q$, as shown in Figure 1.
Since we are mostly interested in the effects of quark loops, there is no need to use a  full-fledged background field method. This two-point function is
\begin{equation}
\Pi^{\mu\nu}(Q)=i(Q^{\mu}Q^{\nu}-Q^2g^{\mu\nu})\Pi(Q)\end{equation}
with
\begin{equation}\label{pint}
\Pi(Q)\approx i \sum_{m_n<\Lambda}\beta_n \int\frac{d^4P}{(2\pi)^4}\frac{1}{P^2+m_n^2}\frac{1}{(P+Q)^2+m_n^2}.\end{equation}
Here, the sum extends over all particles with masses below the ultraviolet cutoff $\Lambda$, and  $\beta_n$ denote the respective beta function coefficients:
\begin{equation}
\beta_n=2(-1)^{F_n}(\lambda_n^2-\frac{1}{12})C_n\, ,
\end{equation}
where $F_k$ is the fermion number, $\lambda_n$ the helicity and $C_n$ the quadratic Casimir in the particle's $SU(3)$ color representation. The momentum dependence of the integral (\ref{pint}) changes at the threshold. In a rough approximation,
\begin{eqnarray}Q\lsim m_t:&\qquad&
\Pi(Q)\approx  \Big[\!\!\sum_{n:\,m_n<m_t}\!\!\!\frac{\beta_n}{(4\pi)^2}\Big]\ln \Big(\frac{Q}{\Lambda}\Big)^2 +\sum_{n:\,\Lambda>m_n\ge m_t}\frac{\beta_n}{(4\pi)^2}\ln \Big(\frac{m_n}{\Lambda}\Big)^2 \nonumber \\[-1mm] &&\label{qsum} \\[1mm] \nonumber
Q\gsim m_t:&\qquad&
\Pi(Q)\approx  \Big[\!\!\sum_{n:\,m_n\le m_t}\!\!\!\frac{\beta_n}{(4\pi)^2}\Big]\ln \Big(\frac{Q}{\Lambda}\Big)^2 +\sum_{n:\,\Lambda>m_n> m_t}\frac{\beta_n}{(4\pi)^2}\ln \Big(\frac{m_n}{\Lambda}\Big)^2
 \end{eqnarray}
Below the threshold, the top quark loop does not participate in the logarithmic running of the coupling constant, which is completely determined by the particle spectrum below $m_t$. However, its contribution ensures a smooth transition to higher energies, where the coupling runs with the beta function coefficient including top. While the full renormalization group beta function determines the cutoff dependence of couplings, the finite threshold effects play an important role in the  evolution of  effective, physical couplings. Thus they are very important for all applications involving extrapolations to high energies, in particular in the framework of unification scenarios.

The fact that even a single particle can produce significant threshold effects is very important for grand unification, but it does not excite imagination in a way like the threshold to higher dimensions, envisaged in some Kaluza-Klein (KK) scenarios beyond the standard model \cite{Antoniadis:1990ew}. When crossing to higher dimensions, including, say,  a circle of radius $R$, one encounters not just one particle, but an infinite tower of KK excitations with masses $m_n=n/R$ labelled by integers $n\ge 0$. For each tower with $\beta_n=\beta_0$, the sums in Eqs.(\ref{qsum}), with the threshold mass $m_t$ replaced by $1/R$, split into
$0\le n<QR$ and $QR<n<\Lambda R$. The latter can be approximated by an integral, giving \cite{Taylor:1988vt}:
\begin{eqnarray}Q\ll 1/R\quad:&\qquad&
\Pi(Q)\approx  \frac{\beta_0}{(4\pi)^2}\Big[\ln (QR)^2 -2({\cal N}-1)\Big]\nonumber \\[-1mm] &&\label{rsum} \\[-3mm] \nonumber
1/R\ll Q\ll\Lambda:&\qquad&
\Pi(Q)\approx  2\frac{\beta_0}{(4\pi)^2}(RQ-{\cal N})\, ,
 \end{eqnarray}
where ${\cal N}=\Lambda R$ is the (large) number of KK excitations below the cutoff $\Lambda$.
Note that the logarithmic running occurs only below the decompactification scale, and is completely determined by the properties of the massless state at the bottom of the tower. Above the threshold, there is a power (linear) running appropriate to non-compact five dimensions. The logarithmic running is something very special to four dimensions -- it is a remnant of infrared divergences that do not appear in higher dimensions. Incidentally,  in order to explain why we live in four dimensions, one needs a mechanism that relies on some special properties of $D=4$, thus infrared divergences are very likely to play a role in such dynamical compactifications \cite{Taylor:1988vt}. Note that the dominant momentum-independent one-loop threshold correction is of order  $\cal O (N)$.

The computations of one-loop threshold effects can be repeated for more general, possibly anisotropic compactifications, always with the same result that 
the radius $R$ which determines the scale of logarithmic running should be understood as the largest length scale characterizing the compact space. Thus the onset of power running occurs as soon as the energies approach the first Kaluza-Klein mass. 

More recently, large radius compactifications became quite a popular element of model building beyond the standard model. Although it is a very attractive possibility, it seems to be incompatible with the existence  of supersymmetric grand unification suggested by the observed values of gauge coupling constants, which is based on the logarithmic running. However, as shown in \cite{Dienes:1998vh}, it is possible that large threshold corrections can also lead to unification,  at lower energy scales, determined by the size of compact dimensions. Here, supersymmetry seems to lose its special appeal, however it is desirable for another reason. A mechanism  based on large one-loop threshold corrections can be reliable only if the higher loop effects are small. Without supersymmetry, there is no reason to expect that this is the case. The common feature of $N=1$ supersymmetric compactifications is that at heavy Kaluza-Klein levels the spectrum as well as interactions are $N=2$ supersymmetric. It is well known that $N=2$ gauge couplings are not renormalized beyond one loop. It can be also shown \cite{Kakushadze:1999bb} that the one-loop threshold corrections are dominant, while the higher loop corrections are suppressed, at least by some powers of the tree-level coupling constants.

\section{Superstring Threshold Corrections}
Threshold corrections appear also in the framework of string theory, which brings two new elements. First, the ultraviolet cutoff is a physical parameter, related to the Regge slope $\alpha'$ that determines the masses of heavy string modes: $\Lambda\approx (\alpha')^{-1/2}$. Thus the cutoff itself becomes a threshold, for the production of heavy string modes. Second, the tree level coupling constants and the radius, as well other geometric quantities characterizing the shape and size of extra dimensions, correspond to vacuum expectation values (VEVs) of certain moduli fields. The moduli parameterize flat directions of the tree-level scalar potential, therefore the determination of their VEVs is a dynamical problem, of  ``moduli stabilization.''
 
The fact that string theory is ultraviolet finite does not prevent gauge couplings from running which, as explained before, is an infrared effect, and can be studied
by using the low energy effective field theory. A more rigorous, formal treatment of threshold corrections is complicated by the fact that only on-shell amplitudes can be computed by using standard string-theoretical techniques. At the same time when Gabriele was using
effective field theory with a string cutoff \cite{Veneziano:1988aj}, Kaplunovsky \cite{Kaplunovsky:1987rp} developed a full-fledged formalism for studying threshold corrections in string theory. Then Dixon, Kaplunovsky and Louis \cite{Dixon:1990pc} studied moduli-dependence of string loop corrections in certain heterotic orbifold compactifications.\footnote{These computations were later extended to more general orbifolds by Mayr and Stieberger \cite{Mayr:1993mq}. More recently, L\"ust and Stieberger \cite{Lust:2003ky} studied gauge threshold corrections in intersecting brane-world models. The formalism for computing threshold corrections to Yukawa couplings has been developed in \cite{Antoniadis:1992pm}.} For any untwisted modulus $T$ upon which the threshold corrections 
$\Delta$ do depend non-trivially, the functional form of this dependence is given by
\begin{equation}\label{dkl}
\Delta=A\cdot\ln\big(|\eta(T)|^4\cdot \makebox{Im}T\big) +T\makebox{-independent terms},
\end{equation}
where $A$ are computable constants determined by the massless spectrum. The Dedekind function is defined by
\begin{equation}
\eta(T)=e^{\pi iT/12}\prod_{n=1}^{\infty}(1-e^{2\pi i nT})\, .
\end{equation}
It is very interesting to compare Eqs.(\ref{rsum}) and (\ref{dkl}). To make it simple, consider a six-dimensional orbifold which is a product of a two-dimensional torus $T^2$ times ``something'' four-dimensional, and that $T^2$ is a product of two circles with radii $R_1$ and $R_2$, respectively. For such compactifications, there exists a modulus parameterizing the volume of $T^2$: Im$T=R_1R_2/\alpha'=(\Lambda R_1)\cdot(\Lambda R_2)$.
Note that Im$T\sim {\cal N}$, measuring also the (approximate) number ${\cal N}$ of KK excitations of $T^2$ with masses below the string cutoff. In the limit of large radii,
${\cal N}\to\infty$, and
\begin{equation}\label{del}
\Delta \;\sim\; -A\cdot \frac{\pi}{3}\,{\cal N}\, ,
\end{equation}
thus the string threshold correction have the same large radius behavior as a generic sum of KK modes (\ref{rsum}), up to a multiplicative constant which is rather ambiguous because it is related to the precise implementation of the mass cutoff on the KK spectrum.
However, the coefficients $A$ are non-zero only for the orbifold sectors with $N=2$ supersymmetry.
Furthermore, according to the non-renormalization theorem proven in \cite{Antoniadis:1991fh}, all
higher loop (genus) corrections are zero exactly, thus the full-fledged string computations are perfectly compatible with the effective field theory analysis. A more precise match between the two formalisms has been discussed in \cite{Gaillard:1992bt}. 

In any closed string theory like the heterotic one, KK modes of each circle are accompanied by strings winding $n$ times around the circle, with masses $m_n=nR/\alpha'$. The spectrum as well as the interactions have a small--large radius symmetry $R
\leftrightarrow \alpha'/R$, which is extended in the orbifold compactifications to a full $T$-duality: $PSL(2,\bf{Z})$ modular invariance generated by $T\to -1/T$ and $T\to T+1$.
The threshold correction (\ref{dkl}) is $PSL(2,\bf{Z})$-invariant. This invariance is realized, however, in quite a non-trivial way. $\Delta$ is the coefficient of
the kinetic energy terms of gauge bosons so its form is restricted by supersymmetry to be
a real part of a holomorphic function of chiral fields. Indeed, at the tree level $g^{-2}=4\,\makebox{Re}S$, where $S$ is the dilaton superfield. The presence of Im$T$ under the logarithm in Eq.(\ref{dkl}), which is necessary for the modular invariance, is in conflict with that property. Thus the string threshold corrections suffer from a ``holomorphic anomaly'' \cite{Derendinger:1991hq,LopesCardoso:1992yd} which is related to the infrared  divergences associated to massless states that {\it cannot} be described in terms of a {\it local\/} effective action \cite{Louis:1991vh}. Since then, holomorphic anomalies play an important role in more formal areas of superstring theory, see {e.g}.\ 
\cite{Bershadsky:1993ta}.

The moduli-dependent threshold corrections have some interesting phenomenological consequences. For example, in some specific orbifold models with the light spectrum below the compactification scale consisting only of the particles belonging to the minimal standard model, a phenomenologically viable gauge coupling unification imposes certain constraints on the modular transformation properties of quark, lepton and Higgs superfields \cite{pheno}.

The fact that gauge (and other) couplings are moduli-dependent may also help is stabilizing the moduli VEVs. In particular, in the context of hidden gaugino condensation mechanism of supersymmetry breaking \cite{nilles}, the scale $\Lambda_{\makebox{\tiny \it SYM}}$ of gaugino condensation is given, in the two-loop approximation, by
\begin{equation}
\Lambda_{\makebox{\tiny \it SYM}}=\mu\, g^{\beta_1/2\beta_0^2} \exp\Big(\frac{-8\pi^2}{\beta_0 g^2}\Big),
\end{equation}
where $\mu$ is the scale at which the gauge coupling constant $g$ is defined, and $\beta_0$, $\beta_1$ are the beta function coefficients of the hidden super Yang-Mills (SYM) sector: $\beta(g)=-\frac{\beta_0}{(4\pi)^2}g^3-\frac{\beta_1}{(4\pi)^4}g^5+\dots$ {}From Gabriele and Shimon's work \cite{Veneziano:1982ah} we know that gaugino condensation can be described in terms of a simple lagrangian for the composite superfield ${\cal W_{\alpha}W^{\alpha}}=\lambda_{\alpha}\lambda^{\alpha}+\dots$,
with the coupling constant promoted to a holomorphic function of the dilaton and moduli superfields \cite{Taylor:1985fz}. In heterotic orbifold compactifications, the moduli-dependence is completely determined by modular invariance \cite{fmtv1,fmtv2}. As an example, consider a pure SYM hidden sector and focus on the dependence of the Veneziano-Yankielowicz lagrangian on three superfields: $\cal W_{\alpha}W^{\alpha}$, the dilaton $S$
and one modulus $T=a+iR^2/\alpha'$, where $R$ is the (common) radius of six compact dimensions and $a$ is the associated axion. One finds \cite{fmtv1} that the condensation occurs at the expected scale
\begin{equation}
|\langle\lambda_{\alpha}\lambda^{\alpha}\rangle |= \Lambda_{\makebox{\tiny \it SYM}}^3
\end{equation}
with the identification:\footnote{This comparison makes use of the fact that $\beta_1/2\beta_0^2=-2/3$ in SYM theory.}
\begin{equation}\mu^2=\frac{1}{2\,\makebox{Im}T}\quad,\qquad \quad \frac{1}{g^2}=4\,\makebox{Re}\big[S+\frac{\beta_0}{(4\pi)^2}\ln\eta (T)\big]\, .
\end{equation}
The above result can be interpreted by saying that the scale $\mu$ is the {\it infrared} cutoff while $g$ is the Wilsonian coupling constant \cite{Kaplunovsky:1995jw} including the non-anomalous part of threshold corrections (\ref{dkl}), due to {\it massive} KK states with $m_n>\mu$ only. This is the coupling that should be used in the effective four-dimensional field theory at energies below the compactification threshold $\mu$, which from the low-energy point of view becomes an {\it ultraviolet} cutoff. Indeed, $g^{-2}$ is a real part of a holomorphic function, as required by supersymmetry. However, it is not modular-invariant because the zero mass modes are excluded from loop integrals.

In order to obtain the moduli superpotential generated by gaugino condensation, one integrates out the  composite field $\cal W_{\alpha}W^{\alpha}$. This leads to the following superpotential:
\begin{equation}\label{superpot}
W=\exp\Big[-\frac{96\pi^2}{\beta_0}\big(S+\frac{\beta_0}{(4\pi)^2}\ln\eta (T)\big)\Big]=
e^{-\frac{96\pi^2}{\beta_0}S}\eta^{-6}(T)\, .
\end{equation}
The above superpotential transforms under the  $PSL(2,\bf{Z})$ modular transformations as a form of weight ${-}3$, which ensures modular invariance of the lagrangian. In fact, its form is determined uniquely by the modular properties and asymptotic behavior, so it can be also derived without necessarily going into details of SYM dynamics. The corresponding scalar potential is modular invariant, therefore it is symmetric under $R
\leftrightarrow \alpha'/R$ and has stationary points at $R^2=\alpha'$ ($T=i$). Since its form  depends on the details of K\"ahler potential, which also receives some nonperturbative corrections,  it is difficult to prove that $T$ and other moduli are stabilized, however there are some indications that this is indeed the case in some models.
As far as the dilaton is concerned, the scalar potential exhibits a ``runaway'' behavior at
$S\to \infty$, driving the model to its trivial zero-coupling limit. This problem can be circumvented if the hidden SYM sector contains a gauge group consisting of several simple subgroup factors. Then the dilaton VEV can be ``locked'' by a ``racetrack'' of the potential \cite{racetr}.

The threshold effects of extra dimensions and the related gaugino condensation mechanism remain as important ingredients of superstring model-building, now including not only heterotic strings, but also D-branes and flux compactifications. They may play a major role in connecting superstring theory to the real world. \vskip 5mm
 
\noindent {\bf Acknowledgments}\\[5mm]
I would like to thank Gabriele for twenty five years of enjoyable collaborations, his friendship, guidance and support. I am looking forward to many future projects,
as exciting and enjoyable as usual. I am also grateful to my collaborators:
Ig{\nolinebreak}natios Antoniadis, Pierre Bin\'etruy, Sergio Ferrara, Mary K. Gaillard, Edi Gava, Zurab Kakushadze, Dieter L\"ust, Nico Magnoli, Narain and Pran Nath, who
worked together with me on the related topics. This work  is supported in part by the U.S.
National Science Foundation Grant PHY-0600304. Any opinions, findings, and conclusions or
recommendations expressed in this material are those of the author and do not necessarily
reflect the views of the National Science Foundation.

%
%
%

%

\begin{thebibliography}{99.}
%
%
%
%
\bibitem{Taylor:1988vt}
  T.~R.~Taylor and G.~Veneziano,
  Phys.\ Lett.\ B {\bf 212}, 147 (1988).
\bibitem{Veneziano:1988aj} G.~Veneziano,
  ``Topics in String Theory,''
CERN-TH-5019/88 (1988).
\bibitem{Antoniadis:1990ew}
  I.~Antoniadis,
  Phys.\ Lett.\ B {\bf 246} (1990) 377.
\bibitem{Dienes:1998vh}
  K.~R.~Dienes, E.~Dudas and T.~Gherghetta,
  Phys.\ Lett.\ B {\bf 436} (1998) 55
  [arXiv:hep-ph/9803466].
\bibitem{Kakushadze:1999bb}
  Z.~Kakushadze and T.~R.~Taylor,
  Nucl.\ Phys.\ B {\bf 562} (1999) 78
  [arXiv:hep-th/9905137].
  %
\bibitem{Kaplunovsky:1987rp}
  V.~S.~Kaplunovsky,
  Nucl.\ Phys.\ B {\bf 307} (1988) 145
  [Erratum-ibid.\ B {\bf 382} (1992) 436]
  [arXiv:hep-th/9205068].
  %
\bibitem{Dixon:1990pc}
  L.~J.~Dixon, V.~Kaplunovsky and J.~Louis,
  Nucl.\ Phys.\ B {\bf 355} (1991) 649.
\bibitem{Mayr:1993mq}
  P.~Mayr and S.~Stieberger,
  Nucl.\ Phys.\ B {\bf 407} (1993) 725
  [arXiv:hep-th/9303017].
\bibitem{Lust:2003ky}
  D.~L\"ust and S.~Stieberger,
  arXiv:hep-th/0302221.
\bibitem{Antoniadis:1992pm}
  I.~Antoniadis, E.~Gava, K.~S.~Narain and T.~R.~Taylor,
  Nucl.\ Phys.\ B {\bf 407} (1993) 706
  [arXiv:hep-th/9212045].
  %
  \bibitem{Antoniadis:1991fh}
  I.~Antoniadis, K.~S.~Narain and T.~R.~Taylor,
  Phys.\ Lett.\ B {\bf 267} (1991) 37.
  \bibitem{Gaillard:1992bt}
  M.~K.~Gaillard and T.~R.~Taylor,
  Nucl.\ Phys.\ B {\bf 381} (1992) 577
  [arXiv:hep-th/9202059].
  \bibitem{Derendinger:1991hq}
  J.~P.~Derendinger, S.~Ferrara, C.~Kounnas and F.~Zwirner,
  Nucl.\ Phys.\ B {\bf 372} (1992) 145.
  \bibitem{LopesCardoso:1992yd}
  G.~Lopes Cardoso and B.~A.~Ovrut,
  Nucl.\ Phys.\ B {\bf 392}, 315 (1993)
  [arXiv:hep-th/9205009].
  \bibitem{Louis:1991vh}
  J.~Louis,
SLAC-PUB-5527 (1991), published in Boston PASCOS 1991:751-765.
\bibitem{Bershadsky:1993ta}
  M.~Bershadsky, S.~Cecotti, H.~Ooguri and C.~Vafa,
  Nucl.\ Phys.\ B {\bf 405} (1993) 279
  [arXiv:hep-th/9302103].
  \bibitem{pheno}L.E.~Ibanez, D.~L\"ust and G.G.~Ross,
Phys.\ Lett.\ B {\bf 272} (1991) 251, hep-th/9109053;
L.E.~Ibanez and D.~L\"ust,
Nucl.\ Phys.\ B {\bf 382}, 305 (1992), hep-th/9202046;
P.~Mayr, H.P.~Nilles and S.~Stieberger,
Phys.\ Lett.\ B {\bf 317} (1993) 53, hep-th/9307171;
H.P.~Nilles and S.~Stieberger,
Phys.\ Lett.\ B {\bf 367} (1996) 126, hep-th/9510009;
Nucl.\ Phys.\ B {\bf 499} (1997) 3, hep-th/9702110.
  \bibitem{nilles}
H.~P.~Nilles,
Phys.\ Lett.\ B {\bf 115} (1982) 193;
S.~Ferrara, L.~Girardello and H.~P.~Nilles,
Phys.\ Lett.\ B {\bf 125} (1983) 457;
M.~Dine, R.~Rohm, N.~Seiberg and E.~Witten,
Phys.\ Lett.\ B {\bf 156} (1985) 55;
C.~Kounnas and M.~Porrati,
Phys.\ Lett.\ B {\bf 191} (1987) 91.
\bibitem{Veneziano:1982ah}
  G.~Veneziano and S.~Yankielowicz,
  Phys.\ Lett.\ B {\bf 113}, 231 (1982).
\bibitem{Taylor:1985fz}
  T.~R.~Taylor,
  Phys.\ Lett.\ B {\bf 164} (1985) 43.
  \bibitem{fmtv1}
S.~Ferrara, N.~Magnoli, T.~R.~Taylor and G.~Veneziano,
Phys.\ Lett.\ B {\bf 245} (1990) 409.
\bibitem{fmtv2}
A.~Font, L.~E.~Ibanez, D.~L\"ust and F.~Quevedo,
Phys.\ Lett.\ B {\bf 245} (1990) 401;
H.~P.~Nilles and M.~Olechowski,
Phys.\ Lett.\ B {\bf 248} (1990) 268;
P.~Bin\'etruy and M.~K.~Gaillard,
Phys.\ Lett.\ B {\bf 253} (1991) 119;
M.~Cvetic, A.~Font, L.~E.~Ibanez, D.~L\"ust and F.~Quevedo,
Nucl.\ Phys.\ B {\bf 361} (1991) 194; D.~L\"ust and T.~R.~Taylor,
  Phys.\ Lett.\ B {\bf 253} (1991) 335; P.~Bin\'etruy, M.~K.~Gaillard and T.~R.~Taylor,
  Nucl.\ Phys.\ B {\bf 455}, 97 (1995)
  [arXiv:hep-th/9504143]; P.~Nath and T.~R.~Taylor,
  Phys.\ Lett.\ B {\bf 548} (2002) 77
  [arXiv:hep-ph/0209282].
\bibitem{Kaplunovsky:1995jw}
  V.~Kaplunovsky and J.~Louis,
  Nucl.\ Phys.\ B {\bf 444} (1995) 191
  [arXiv:hep-th/9502077].
\bibitem{racetr}
  N.~V.~Krasnikov,
  Phys.\ Lett.\ B {\bf 193} (1987) 37;
  L.J. Dixon, SLAC-PUB-5229 (1990),
published in DPF Conf.1990:811-822;
  T.~R.~Taylor,
  Phys.\ Lett.\ B {\bf 252}, 59 (1990).
\end{thebibliography}
%



\printindex
\end{document}